\documentclass[fleqn,usenatbib]{mnras}

\usepackage{newtxtext,newtxmath}
\usepackage[T1]{fontenc}
\usepackage{subfig} %for figure subplot
\usepackage{graphicx}
\usepackage[flushleft]{threeparttable}
\usepackage{xspace}

\newcommand{\hst}{{\it HST}}
\newcommand{\hc}{$H_0$}

\newcommand{\mbh}{$\mathcal M_{\rm BH}$}
\newcommand{\lhost}{$L_{\rm host}$}

\newcommand{\Hb}{H$_{\beta}$}
\newcommand{\sersic}{S\'ersic}
\newcommand{\lenstronomy}{{\sc Lenstronomy}}
\newcommand{\reff}{{$R_{\mathrm{eff}}$}}

\newcommand{\kms}{\ifmmode{\,\rm{km}\, \rm{s}^{-1}}\else{$\,$km$\,$s$^{-1}$}\fi}

\newcommand{\mstar}{{$M_*$}}
\newcommand{\Mgii}{Mg$_{\rm II}$}
\newcommand{\Civ}{C$_{\rm IV}$}
\DeclareRobustCommand{\VAN}[3]{#2}
\let\VANthebibliography\thebibliography
\def\thebibliography{\DeclareRobustCommand{\VAN}[3]{##3}\VANthebibliography}

\usepackage{graphicx}	% Including figure files
\usepackage{amsmath}	% Advanced maths commands

\usepackage{amssymb}	% Extra maths symbols

\title[Mass correlations of lensed AGN hosts]{Testing the Evolution of the Correlations between Supermassive Black Holes and their Host Galaxies using Eight Strongly Lensed Quasars}

\author[X. Ding et al.]{
Xuheng Ding,$^{1}$\thanks{E-mail: dxh@astro.ucla.edu}
Tommaso Treu,$^{1}$
Simon Birrer,$^{2}$
Adriano Agnello,$^{3}$\newauthor
Dominique Sluse,$^{4}$
Chris Fassnacht,$^{5}$
Matthew W. Auger,$^{6}$
Kenneth C. Wong,$^{7}$ \newauthor
Sherry H. Suyu,$^{8,9,10}$
Takahiro Morishita,$^{11}$
Cristian E. Rusu,$^{12,13,14}$
Aymeric Galan,$^{15}$
\\
\\
$^{1}$Department of Physics and Astronomy, University of California, Los Angeles, CA, 90095-1547, USA\\
$^{2}$Kavli Institute for Particle Astrophysics and Cosmology and Department of Physics, Stanford University, Stanford, CA 94305, USA\\
$^{3}$European Southern Observatory, Karl-Schwarzschild-Strasse 2, 85748 Garching, Germany\\
$^{4}$STAR Institute, Quartier Agora - All\'ee du six Ao\^ut, 19c B-4000 Li\`ege, Belgium\\
$^{5}$Department of Physics and Astronomy, University of California, Davis, CA 95616, USA\\
$^{6}$Institute of Astronomy, University of Cambridge, Madingley Road, Cambridge CB3 0HA, UK\\
$^{7}$Kavli IPMU (WPI), UTIAS, The University of Tokyo, Kashiwa, Chiba 277-8583, Japan\\
$^{8}$Max-Planck-Institut f{\"u}r Astrophysik, Karl-Schwarzschild-Str.~1, 85748 Garching, Germany\\
$^{9}$Physik-Department, Technische Universit\"at M\"unchen, James-Franck-Stra\ss{}e~1, 85748 Garching, Germany\\
$^{10}$Academia Sinica Institute of Astronomy and Astrophysics (ASIAA), 11F of ASMAB, No.1, Section 4, Roosevelt Road, Taipei 10617, Taiwan\\
$^{11}$Space Telescope Science Institute, 3700 San Martin Drive, Baltimore, MD 21218, USA\\
$^{12}$National Astronomical Observatory of Japan, 2-21-1 Osawa, Mitaka, Tokyo 181-8588, Japan\\
$^{13}$Subaru Telescope, National Astronomical Observatory of Japan, 650 N Aohoku Pl, Hilo, HI 96720\\
$^{14}$Department of Physics, University of California, Davis, 1 Shields Avenue, Davis, CA 95616, USA\\
$^{15}$Institute of Physics, Laboratory of Astrophysics, Ecole Polytechnique 
F\'ed\'erale de Lausanne (EPFL), Observatoire de Sauverny, 1290 Versoix, 
Switzerland
}

\date{Accepted XXX. Received YYY; in original form ZZZ}

\pubyear{2020}

\begin{document}
\label{firstpage}
\pagerange{\pageref{firstpage}--\pageref{lastpage}}
\maketitle

\begin{abstract}
One of the main challenges in using high redshift active galactic nuclei to study the correlations between the mass of the supermassive Black Hole (\mbh) and the properties of their active host galaxies is instrumental resolution. Strong lensing magnification effectively increases instrumental resolution and thus helps to address this challenge. In this work, we study eight strongly lensed active galactic nuclei (AGN) with deep {\it Hubble Space Telescope} (\hst) imaging, using the lens modelling code \lenstronomy\ to reconstruct the image of the source. Using the reconstructed brightness of the host galaxy, we infer the host galaxy stellar mass based on stellar population models.  \mbh\ are estimated from broad emission lines using standard methods. Our results are in good agreement with recent work based on non-lensed AGN, demonstrating the potential of using strongly lensed AGNs to extend the study of the correlations to higher redshifts.
At the moment, the sample size of lensed AGN is small and thus they provide mostly a consistency check on systematic errors related to resolution for the non-lensed AGN. However, the number of known lensed AGN is expected to increase dramatically in the next few years, through dedicated searches in ground and space based wide field surveys, and they may become a key diagnostic of black hole and galaxy co-evolution. 
\end{abstract}


%%%%%%%%%%%%%%%%%%%%%%%%%%%%%%%%%%%%%%%%%%%%%%%%%%

%%%%%%%%%%%%%%%%% BODY OF PAPER %%%%%%%%%%%%%%%%%%
\section{Introduction}
The tight correlations between the masses (\mbh) of supermassive black holes (BHs) and their host galaxies properties including stellar mass (\mstar), stellar velocity dispersion ($\sigma_*$) and luminosity (\lhost), known as scaling relations, are usually considered as a result of their co-evolution ~\citep[e.g.,][]{Mag++98, F+M00, Geb++01b, M+H03, Gul++09,Beifi2012, H+R04, Gra++2011}. Cosmological simulations of structure formation are able to reproduce the local tight relation based on the physical mechanism by invoking active galactic nucleus (AGN) feedback as the physical connection~\citep{Springel2005, Hopkins2008, Matteo2008, DeG++15} or having them share a common gas supply~\citep{Cen2015, Menci2016}. However, it has also been suggested that the correlations arise statistically, without any physical coupling, as a result of stochastic mergers~\citep{Peng2007, Jahnke2011, Hirschmann2010}. 

A powerful way to understand the origin of the correlations is to study them as a function of redshift, determining how and when they emerge and evolve over cosmic time~\citep[e.g.,][]{TMB04,Sal++06,Woo++06, Jah++09,SS13,Sun2015, Park15}. Recently, based on a sample of 32 X-ray-selected type-1 AGNs in deep survey fields, \citet{Ding2020a} (hereafter, D20) measured the scaling relations in the redshift range $1.2<z<1.7$ using multi-color Wide Field Camera 3 (WFC3) \hst\ imaging. Combining the new sample with published samples in both local and intermediate (i.e., $0.35<z\lesssim1.2$) redshift ranges, D20 strengthen the support for an evolution scenario in which the growth of BHs evolution predates that of  the bulge component of the host galaxy. In a follow-up paper, \citet{Ding2020b} compared the D20 measurements to the predictions by the numerical simulations including the hydrodynamic simulation MassiveBlackII~\citep{Khandai2015} and a semi-analytic model~\citep{Menci2014}. The observed tightness of the  scaling relations at high redshift is consistent with the hypothesis that AGN feedback drives the scaling correlations, and disfavors the hypothesis of the correlations being purely stochastic in nature. 
% , a larger sample at the higher redshift range would be useful to increase the resolution of the evolution and confirm these hypothesises.

As the samples of AGN grow to increase statistical power, it is very important to make sure that systematics do not dominate the error budget. One of the main potential sources of systematics is the finite resolution of \hst\ images. Even with modern techniques, AGN hosts remain barely resolved at \hst\ resolution, and therefore it would be very useful to verify the results at higher resolution. This is the goal of this work.

It has been long recognized that lensed AGNs, by virtue of magnification, can provide unique insights into the scaling relations the distant universe~\citep{Peng2006}, provided that the lens modelling could be accurately derived. Aiming at verifying the fidelity of the modern lens modelling technique, \citet{Ding2017a} carried out extensive and realistic simulations tests based on the deep \hst\ observations for a sample of eight lens systems in the \hc\ Lenses in COSMOGRAIL's Wellspring \citep[H0LiCOW\footnote{\url{http://www.h0licow.org/}},][]{Suyu2017} collaboration. They confirm that the reconstruction of the lensed host galaxy properties can be recovered with better precision and accuracy than the typical \mbh\ uncertainty. Then, \citet{Ding2017b} applied the advanced techniques to two strongly lensed systems analyzed by the H0LiCOW collaboration~\citep{Suyu2013, Wong2017} to study their \mbh-\lhost\ relations and obtained consistent scaling relations compared with the samples from the literature.

In this work, we expand the measurements of the \mbh-\mstar\ relation using the full sample of eight lensed AGN introduced by \citet{Ding2017a}.  In order to take advantage of the excellent quality of the data, we develop an independent approach to achieve a one-step inference of the host galaxy photometry from the eight lensed AGNs. We adopt a set of extended modelling choices to estimate the host property uncertainty level and ensure the accuracy of our measurements. We compare the inference of our new measurements to the ones that have been modelled by the H0LiCOW collaboration to make a cross-check. To obtain an accurate inference of stellar mass, we utilize the multi-band imaging data taken with the \hst\ to obtain the color information for 3/8 of our targets. We assume a typical stellar population for the rest 5/8 of the sample, consistent with D20. Furthermore, we adopt a class of self-consistent recipes to recalibrate the \mbh\ of our sample, in a manner consistent with D20. Given the similar redshift range, the high data quality, and consistent techniques, this sample of lensed AGN provides an excellent validation of the D20 results.

This paper is structured as follows. In Section~\ref{sec:sample_select}, we introduce the sample including their imaging data and the BH masses. In Section~\ref{sec:lens_model}, we describe our approach designed to infer the lens models and reconstruct the host galaxy. We use the inferred photometry to derive the stellar mass and the scaling relations and combine with D20 sample to study the evolution in Section~\ref{sec:result}. Discussion and conclusions are drawn in Sections~\ref{sec:con}. Throughout this paper, we adopt a standard concordance cosmology with parameters adopted as $H_0= 70$ km s$^{-1}$ Mpc$^{-1}$, $\Omega{_m} = 0.30$, and $\Omega{_\Lambda} = 0.70$, to compute the luminosity distance and estimate the host absolute brightness. Magnitudes are presented in the AB system. A Chabrier initial mass function (IMF) is employed for the sample, to be consistent with D20.

\section{Sample Selection and Black Hole Mass Estimates}\label{sec:sample_select}
We adopt eight lens systems from the H0LiCOW collaboration including HE0435$-$1223, RXJ1131$-$1231, WFI2033$-$4723, HE1104$-$1805, SDSS1206$+$4332, SDSS0246$-$0825, HE0047$-$1756 and HS2209$+$1914. We refer to~\citet{Suyu2017, Ding2017a} for the descriptions of these data. For conciseness, in the rest of this paper, we abbreviate each lens name to four digits (e.g., HE0435$-$1223 to HE0435). 

Based on the observational data of these eight systems, \citet{Ding2017a} performed an extensive and realistic simulation exercise using the \hst\ image data and confirmed that the source reconstruction using the lens modelling technique is trustworthy. We summarize the information for the eight systems, including their redshift, data properties, and references in Table~\ref{data_set}.
Besides the imaging data shown in this table, we also analyzed the multi-band \hst\ imaging data to derive the host color information for 3/8 systems. As we show in Section~\ref{sec:mstar}, we use the color information to fit for the best stellar population to improve the accuracy of the estimate of \mstar. 

The sample of lensed AGN is too limited in size to constrain evolution by itself. Therefore, we use it primarily to verify the results of D20.  The D20 sample includes 32 AGN measurements in the redshift range of $1.2<z<1.9$. They also collected 59 intermediate redshift (i.e., $0.35<z\lesssim1.2$) AGN measurements~\citep{Bennert11, SS13, Cisternas2011} and 55 local (i.e., $z\lesssim0.007$) measurements~\citep{Bennert++2011, H+R04}. It is worth noting that they are so far the largest \hst\ imaging AGN sample with redshift range up to $z\sim1.9$. 
 
\begin{table*}
\centering
%\begin{threeparttable}
\caption{Summary of lensed AGN observational details.}\label{data_set}
\resizebox{15cm}{!}{
     \begin{tabular}{ccccccccc}
        \hline
Object ID & $z_s$ & camera & Filter & exposure & Program ID & PI & pixel scale & References \\
 & & & & time (s) &&& (drizzled) & \\ \hline
HE0435$-$1223 & 1.69 & WFC3-IR & F160W & 9340 & 12889 & S. H. Suyu & $0\farcs{}08$ & (1), (2)\\
RXJ1131$-$1231 & 0.65 & ACS & F814W & 1980 & 9744 & C.S. Kochanek & $0\farcs{}05$ &(3), (4)\\
WFI2033$-$4723 & 1.66 & WFC3-IR & F160W & 26257 & 12889 & S. H. Suyu & $0\farcs{}08$ & (5), (2) \\
SDSS1206$+$4332 & 1.79 & WFC3-IR & F160W & 8457 & 14254 & T. Treu & $0\farcs{}08$ & (6), (7)\\
HE1104$-$1805 & 2.32 & WFC3-IR & F160W & 14698 & 12889 & S. H. Suyu  & $0\farcs{}08$ & (8), (9)\\
SDSS0246$-$0825 & 1.68 & WFC3-UVIS & F814W & 8481 & 14254 & T. Treu & $0\farcs{}03$& (10), (11) \\
HS2209$+$1914$^a$  & 1.07 & WFC3-UVIS & F814W & 9696 + 4542 & 14254 & T. Treu & $0\farcs{}03$ & (12), (13) \\
HE0047$-$1756 & 1.66 & WFC3-UVIS & F814W & 9712 & 14254 & T. Treu & $0\farcs{}03$ & (14), (15) \\
        \hline
     \end{tabular}
    }
\begin{tablenotes}
      \small
      \item Note: $-$ The observational information is also given by~\citet{Ding2017a}.
      \item $^a$: HS2209, was visited by \hst\ twice ({\it vis05} and {\it vis06}) at different orientations. The exposure time is thus given separately.
      \item References:$-$ (1) \citet{HE0435_discover}; (2) \citet{Sluse2012}; (3) \citet{1131_discover}; (4) \citet{1131_redshift}; (5) \citet{2033_discover}; (6) \citet{1206_discover}; (7) \citet{Eul++13}; (8) \citet{HE1104_discover}; (9) \citet{1104_d_redshift}; (10) \citet{0246_discover}; (11) \citet{0246_redshift}; (12) \citet{2209_discover};  (13) \citet{2209_more}; (14) \citet{0047_discover}; (15) \citet{Ofe++06};
    
\end{tablenotes}  
%\end{threeparttable}
\end{table*}

To ensure the consistency of \mbh\ estimates based on different broad lines, we adopt the following set of self-consistent recipes following D20:
\begin{eqnarray}
\label{recipe}
\log \left(\frac{\mathcal M_{\rm BH}}{M_{\odot}}\right)&~=~& a+b \log \left(\frac{ \rm L _{\lambda_{line}}}{10^{44}{\rm erg~s^{-1}}}\right) \nonumber\\
&~+~& 2 \log \left(\frac{\rm FWHM(line)}{1000 ~{\rm km~s^{-1}}}\right) , 
\end {eqnarray}
where $\lambda_{\rm line}$ is the reference wavelength of the local continuum luminosities for different emission lines. The following values are adopted:
$a$\{\Civ, \Mgii, \Hb\}=\{6.322, 6.623, 6.910\},
$b$\{\Civ, \Mgii, \Hb\}=\{0.53, 0.47, 0.50\},
$\lambda_{\rm line}$\{\Civ, \Mgii, \Hb\}=\{1350, 3000, 5100\}~(\AA).
The broad line properties of our samples are adopted from the literature, with a few corrections/exceptions. % \citep{Sluse2012, Peng2006, Shen2011}.
The line properties of SDSS1206 have been inferred by~\citet{Shen2011}. However, this system was investigated as a non-lensed AGN and the lensing magnification on the intrinsic continuum luminosity was not considered. We follow~\citet{Birrer2019} and apply the same magnification correction on the $\log(L_\lambda)$ in this work. For the rest of the lens sample except HS2209, their broad line properties are adopted from the literature~\citep{Sluse2012, Peng2006} taking into account lensing magnification. For HS2209, the spectrum was observed at the Keck-II Telescope in September 2015. We derived the line properties from the spectrum using the same approach as~\citet{Sluse2012}, based on \Mgii.
The \mbh\ measurements, together with the properties of the broad-line, are listed in Table~\ref{mbh}. 
The uncertainty of the \mbh\ is estimated to be $0.4$~dex.
We note that the magnification correction for these systems are not fully self-consistent, in the sense that the magnification is slightly different for the host galaxy light and the black holes mass. This difference could  in principle introduce some systematics on the correlations. However, these errors are smaller than the calibration error on \mbh (0.4 dex), given that the virial relations depend on roughly the square root of the $\log(L_\lambda)$.

\begin{table}
\centering
%\begin{threeparttable}
\caption{Summary of \mbh\ estimates, based on equation~\eqref{recipe}.}\label{mbh}
\resizebox{8.5cm}{!}{
     \begin{tabular}{cccccc}
        \hline
Object ID & Line(s) & FWHM & log($L_\lambda$) & $\log$\mbh & ref.  \\
& & (\kms) & (${\rm erg~s^{-1}}$) & (M$_{\odot}$) \\ \hline
HE0435 & \Mgii & 4930 & 45.14 & 8.54 & (1) \\
RXJ1131 & \Mgii/\Hb & 5630/4545 & 44.29/44.02 & 8.26/8.23 & (1)\\
WFI2033 & \Mgii & 3960 & 45.19 & 8.38 & (1) \\
SDSS1206 & \Mgii & 5632 & 45.01 & 8.60 & (2) \\
HE1104 & \Civ & 6004 & 46.18 & 9.03 & (3) \\
SDSS0246 & \Mgii & 3700 & 45.19 & 8.32 & (1) \\
HS2209 & \Mgii & 3245 & 45.71 & 8.45 & here  \\
HE0047 & \Mgii & 4145 & 45.59 & 8.60 & (1) \\
        \hline
     \end{tabular}
    }
\begin{tablenotes}
\small
\item Note: $-$ The broad line properties.\\ %For HS2209, the properties are derived in this work using the same approach as~\citet{Sluse2012}.
Reference: (1)~\citet{Sluse2012}, (2)~\citet{Shen2011}, (3)~\citet{Peng2006}.\\
%$^a$: For SDSS1206, we follow~\citet{Birrer2019} and correct the AGN continuum luminosity $\log(L_\lambda$) by taking account the magnification effect {\bf [TT: how about the others?].}
\end{tablenotes}  
%\end{threeparttable}
\end{table}

\section{surface photometry inference}\label{sec:lens_model}
In this section, we describe how we derived surface photometry of the lensed galaxies taking lensing effects into account. Lens models of 4/8 systems, namely HE0435~\citep{Wong2017}, RXJ1131~\citep{Suyu2013}, WFI2033~\citep{Rusu2019} and SDSS1206~\citep{Birrer2019} have been published by the H0LiCOW project. The goal of those models was strong lens time-delay cosmography~\citep{Refsdal1966, Treu2016}, and the reconstruction of the host galaxy light (via pixellated distribution~\citep{Suy++06} or shapelets~\citep{Refregier2003}) was a byproduct. \citet{Ding2017b} used those reconstructions for two of the systems (HE0435 and RXJ1131) and then fitted a simply parametrized surface brightness profile to the reconstruction to measure the host properties as the non-lensed AGN case.
% mimic what is done in non-lensed AGN.

In this work, in order to reproduce more closely and uniformly what is
done in non-lensed AGN, we develop a strategy to obtain a one-step
measurement of the host galaxy light described by a \sersic\ surface
brightness profile. For the 4/8 systems already studied by the H0LiCOW
project, we will make a comparison to characterize systematic
uncertainties related to the modelling techniques.

\subsection{Data Preparation and Modelling Setup}\label{sec:Modelling}

We follow the standard procedure as described by D20 to prepare the fitting ingredients including the lensing imaging data, noise level map and the PSF information. The imaging data are first drizzled to a higher resolution with a Gaussian kernel; the adopted resolutions are listed in Table~\ref{data_set}. We then adopt the {\sc Photutils}~\citep{photutils} Python package to model the global background light in 2-D, based on the \texttt{SExtractor} algorithm. We remove the background light and cut the clear image data into postage stamp at a suitable size. We draw the image mask for each system to define the region in which the pixels will be used to calculate the likelihood; see the top-middle panel in each subplot in Figure~\ref{fig:image_inference}.

We carry out a forward modelling process to simultaneously constrain the lens model, subtract the central AGN light, and infer the photometry of the host galaxy. For any extended objects, including the lensing galaxy and source galaxy, we assume their surface brightness can be described by the 2-D elliptical \sersic\ profile. We start with a single \sersic\ profile and consider using two \sersic\ if any significant residual indicates multiple components. For a single \sersic\ profile, we set the prior of the \sersic\ index value $n$ between $[0.5-5.0]$ to avoid unphysical results\footnote{It has been shown in~\citet{Ding2017a} that choosing this prior for \sersic\ index $n$ yields unbiased host magnitude inferences.}. The bright nucleus is unresolved and modelled by a scaled point spread function (PSF) in the image plane. We also impose that the AGN and its host galaxy have the same center. Following standard practice that has been shown to produce good models \citep[e.g.,][and references therein]{Treu2010}, we adopt elliptical power-law density profiles to define the surface mass density of the deflector, with an external shear.

We employ the imaging modelling tool \lenstronomy\footnote{\url{https://github.com/sibirrer/lenstronomy}}~\citep{lenstronomy} to perform the fitting task, using the ``particle swarm optimizer'' mode. Building on D20, we adopt a set of  modelling choices and fit each system multiple times. Then, we do a statistical analysis of the measurements, and apply a weighting algorithm to derive the final inference and estimate the uncertainty level. The modelling choices that we consider include the follows.
\begin{enumerate}
\item Following common practice, we select all the bright, isolated stars across the image frame of targets to define the PSF. Each selected star is considered as an initial PSF guess for the fit.
\item The central pixels of the AGNs are very bright and can be affected by large systematic errors during the interpolation of the subsampled PSF. To avoid overfitting the noise, we adopt two different modelling options: 1)~manually boosting the noise estimate in the central area to effectively infinite ({\it noise boost}); 2)~performing the iterative PSF estimation as introduced by~\citet{Chen2016, Birrer2019} ({\it PSF iteration}).
\item To calculate the ray tracing under a higher resolution grid relative to the pixel sizes in the image plane, we choose to oversample the model by a factor  $2\times2$ or $3\times3$ pixel$^{-1}$.
\item Using the lensing imaging alone, it is difficult to constrain the power-law slope, especially with our simplified model of the host light distribution. To mitigate the overfitting of the slope value, we repeat the fit for three values $[1.9, 2.0, 2.1]$ that are meant to bracket the range observed in lens galaxies \citep[e.g.,][]{Koopmans2006, Auger2010}.
\end{enumerate}

In general, for one lens system with a number of $N$ initial PSF guesses, we perform in total $N\times12$ (i.e., $2$ by (ii) $\times~2$ by (iii) $\times~3$ by (iv)) fits. After all fits are completed, we rank their performance based on their best-fitted $\chi^2$ value. Since there is no evidence of better performance for the options between {\it noise boost} and {\it PSF iteration}, we selected the top-4 fittings from each of them and then combined their best-fit results based on the weighting algorithm introduced by D20. The degrees of freedom are the same for each of the model, ensuring that the weighting scheme is internally self-consistent. The weights are defined by:
\begin{eqnarray}
\label{eq:weights}
w_i = {\rm exp} \big(- \alpha \frac{ (\chi_i ^2 - \chi_{\rm best} ^2 )}{2 \chi_{\rm best} ^2} \big),
\end{eqnarray} 
where the $\alpha$ is an inflation parameter\footnote{Defining $\alpha$ as the {\it inflation} parameter it is assumed that it is larger than ~$1$, so as to err on the side of caution and include more choices that would be allowed strictly by statistical noise considerations.} so that when $i=4$:
\begin{eqnarray}
\label{eq:alpha}
\alpha \frac{ \chi_{i=4} ^2 - \chi_{\rm best} ^2 }{2 \chi_{\rm best} ^2} = 2.
\end{eqnarray} 
Then, the results for {\it noise boost} and {\it PSF iteration} are combined equally to derive the value of host properties and the root-mean-square error (i.e., $\sigma$ level) based on the weights of the $4+4$ (i.e., {\it noise boost} and {\it PSF iteration}) options by:
\begin{eqnarray}
\label{eq:infer_value}
\bar{x}  =  \frac{  \sum_{i=1}^{N}   x_i * w_i  }{\Sigma w_i} ,
\end{eqnarray} 
\begin{eqnarray}
\label{eq:infer_scatter}
\sigma =   \sqrt{ \frac{  \sum_{i=1}^{N}   (x_i -  \bar{x} ) ^2 * w_i  }{\Sigma w_i} }.
\end{eqnarray} 

Our approach uses the relative goodness of fit and ensures that at least 8 sets of best fits are used to estimate the range of systematic uncertainties. Note that the slope of the mass profile of the deflector is fixed in each fit, and that the statistical uncertainty is much smaller than the systematic one. 

The inferred photometric properties of the host galaxy for all the eight systems are listed in Table~\ref{tab:host_measure}, (2)$-$(6) columns. Detailed information on the fitting process for each system is given in Appendix~\ref{sec:photometry}.

\begin{table*}
\renewcommand{\arraystretch}{1.25}
\centering
  \begin{threeparttable}
\caption{Summary of lensed AGN host inference.}\label{tab:host_measure}    
%\resizebox{12cm}{!}{
     \begin{tabular}{cccccccc}
\hline     
(1) & (2) & (3) & (4) & (5) & (6) & (7) & (8) \\    
Object ID & intrinsic magnitude & magnitude & Host Flux Ratio (to Total) & \reff\ & \sersic\ $n$ & stellar population & $\log (M_{*}$)  \\
 & (source plane) & (image plane) & ( $\%$, Total = Host + AGN) & (arcsec) & & age (Gyr) & (M$_{\odot}$) \\ \hline
HE0435 & $21.49\substack{+0.40\\-0.29}$ & $18.58\substack{+0.30\\-0.23}$ & $36.0\pm11.1$ & $0.28\pm0.02$ & $2.71\pm0.20$ & $1.50$ & $10.91$ \\
RXJ1131$_{\rm bulge}$ & $21.80\substack{+0.23\\-0.19}$ & $18.70\substack{+0.07\\-0.06}$ & $7.1\pm1.4$ & $0.13\pm0.02$ & fix to 4 & $3.00$ & $10.39$ \\
RXJ1131$_{\rm disk}$ & $19.33\substack{+0.17\\-0.15}$ & $17.14\substack{+0.08\\-0.07}$ & $69.2\pm10.1$ & $0.90\pm0.06$ & fix to 1 & $1.50$ & $11.08$ \\
WFI2033 & $21.78\substack{+0.28\\-0.23}$ & $19.07\substack{+0.35\\-0.26}$ & $19.6\pm4.5$ & $0.28\pm0.02$ & $0.52\pm0.01$ & $0.625$ & $10.51$ \\
SDSS1206 & $21.31\substack{+0.23\\-0.19}$ & $18.30\substack{+0.05\\-0.05}$ & $33.3\pm6.4$ & $0.11\pm0.02$ & $4.57\pm0.53$ & $0.625$ & $10.77$ \\
HE1104 & $21.25\substack{+0.16\\-0.14}$ & $19.16\substack{+0.02\\-0.02}$ & $14.0\pm2.0$ & $0.27\pm0.02$ & $1.05\pm0.04$ & $0.625$ & $11.05$ \\
SDSS0246 & $23.44\substack{+0.28\\-0.22}$ & $20.85\substack{+0.08\\-0.07}$ & $4.0\pm0.9$ & $0.44\pm0.08$ & $4.96\pm0.08$ & $0.625$ & $10.75$ \\
HS2209 & $20.72\substack{+0.26\\-0.21}$ & $19.20\substack{+0.04\\-0.04}$ & $12.5\pm2.7$ & $1.96\pm1.28$ & $3.15\pm1.40$ & $1.00$ & $11.04$ \\
HE0047 & $22.92\substack{+0.48\\-0.33}$ & $20.37\substack{+0.20\\-0.17}$ & $2.3\pm0.8$ & $0.32\pm0.15$ & $4.18\pm0.75$ & $0.625$ & $10.91$ \\
\hline
\end{tabular}
%}
\begin{tablenotes}
      \small
      \item Note: $-$ Inference of the host galaxy properties. Column (2)-(6): photometry derived using the imaging data is listed Table~\ref{data_set}. Column (7): adopted age of stellar population with solar metallicity. Column (8): inferred stellar mass. The uncertainty of \mstar\ is estimated to be 0.2 dex.
\end{tablenotes}    
\end{threeparttable}
\end{table*}

\section{Results}\label{sec:result}
In this section, we describe the approaches and assumptions used to estimate the stellar populations in our sample. Then, we adopt stellar population templates to derive the stellar mass. For this step, we infer the color information for the multi-band imaging data taken with \hst.
We study the \mbh-\mstar\ relations and compare our measurements with ones taken from the literature.

\subsection{Stellar population and mass}\label{sec:mstar}
%[] Color inference for the first three cases to decide their stellar template. 
Besides the imaging data analyzed in the last section, some of the systems have also been imaged by \hst\ through other bands, providing color information. Given that we have used the highest signal-to-noise ratio data for our primary models,  the analysis of the other bands has lower fidelity and we use it only to infer the colors so as to assist in the estimation of the stellar mass, i.e., Table~\ref{tab:host_measure} column (2).

\begin{figure*}
\centering
\subfloat[HE0435 F555W]{\includegraphics[trim = 15mm 0mm 15mm 5mm, clip, width=0.9\textwidth]{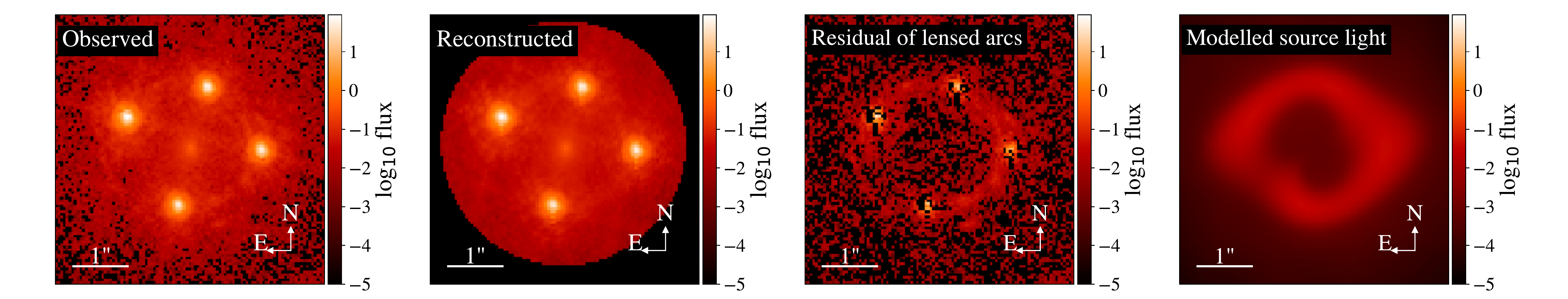}}\\
\subfloat[HE0435 F814W]{\includegraphics[trim = 15mm 5mm 15mm 0mm, clip, width=0.9\textwidth]{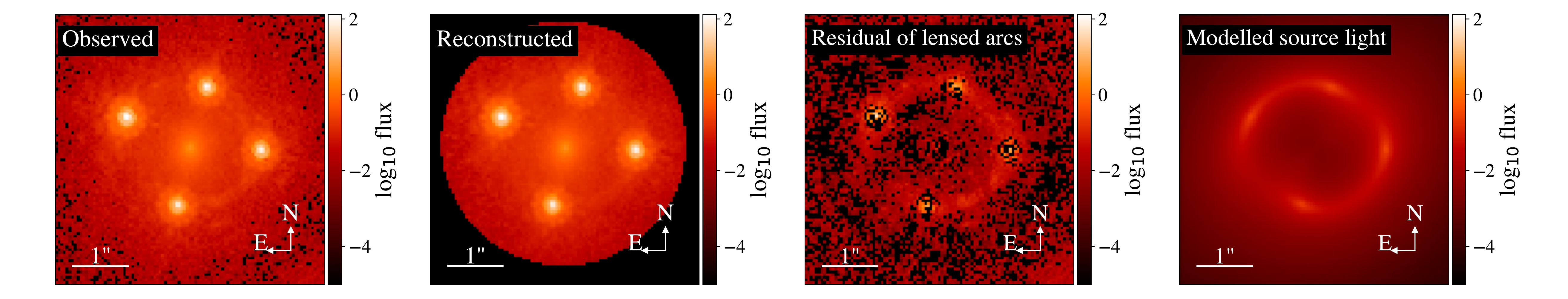}}\\
\subfloat[Host galaxy stellar template inference]{\includegraphics[trim = 10mm 0mm 20mm 15mm, clip, width=0.6\textwidth]{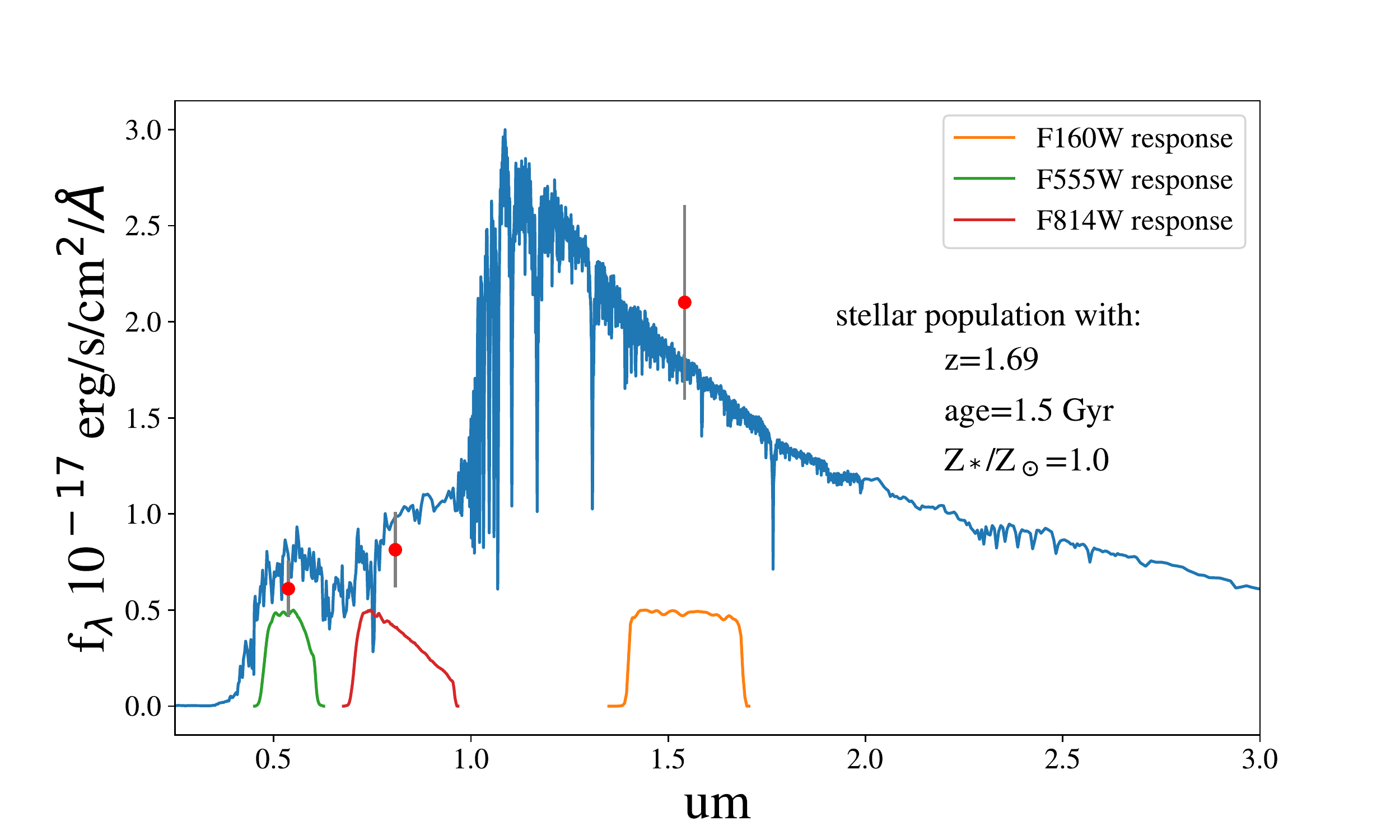}}\\
\caption{\label{fig:app_HE0435} 
Illustrations of the inference of HE0435 using the multi-band data. Top two panels: best-fit model of the lensed arcs shown as Figure~\ref{fig:image_inference}. Bottom panel: SED inference. The red points with error bars represent the inferred image plane host flux in the three bands.}
\end{figure*}

{\bf HE0435} ~
%In addition to the F160W data, this system has been observed by \hst\ through filters F814W and F555W. The multi-band imaging data provides us with the opportunity to estimate the host color and improve the stellar mass estimate. We adopted the lensing modelling approach described in Section~\ref{sec:photometry} to infer the host photometry through the F814W and F555W bands. Based on the three band host magnitudes, we find that a 1.5~Gyr age and solar metallicity stellar population provides a good match to the color.
In addition to WFC3/F160W, HE0435 has been observed through bands ACS/F814W and ACS/F555W (GO-9744; PI: C. S. Kochanek). Our aim is to derive the brightness of the host galaxy through all the three bands to investigate the color and stellar population in the image plane. We use the modelling approach introduced in Section~\ref{sec:photometry} to perform this inference for the F814W and F555W bands. To save computer time, we only use the {\it PSF iteration} approach. We also fix the lens mass slope value to 1.9 since it is closer to the inference by~\citet[][i.e., $\gamma\sim1.93$]{Wong2017}. The inference of the fittings are shown in Figure~\ref{fig:app_HE0435}-(a), (b). Having obtained the magnitude of the lensed host in the three bands, we use the {\sc Gsf} package\footnote{\url{https://github.com/mtakahiro/gsf}}~\citep{Morishita2019} to perform the SED fitting. A range of ages (up to 3.0 Gyr) is used in this fit with a constant SFR and a flexible form for star formation histories. Note that there is a well-known degeneracy between age and metallicity; however, this degeneracy has little effect on the \mstar\ inference~\citep{Bell2001}. In this work, we fix the metallicity to infer the age. In the end, a stellar population with 1.5~Gyr of age and solar metallicity provides the best fit to the colors, as shown in Figure~\ref{fig:app_HE0435}-(c), although this choice is by no means unique. This stellar population is used to estimate the host stellar mass. 

{\bf RXJ1131} ~ The host galaxy of RXJ1131 is lensed to a very extended arc in the image plane. The spectral energy distribution of the arc can be directly inferred in the image plane, since lensing is achromatic. Based on the \hst\ imaging data through the three filters F814W and F555W and F160W, we adopt the SED estimated by~\citet{Ding2017b} with stellar populations of 3 Gyr and 1.5 Gyr (solar metallicity) for its bulge and disk, respectively. We refer the interested reader to that paper for more details.

\begin{figure*}
\centering
\subfloat[WFI2033 F125W]{\includegraphics[trim = 75mm 12mm 45mm 12mm, clip, width=0.9\textwidth]{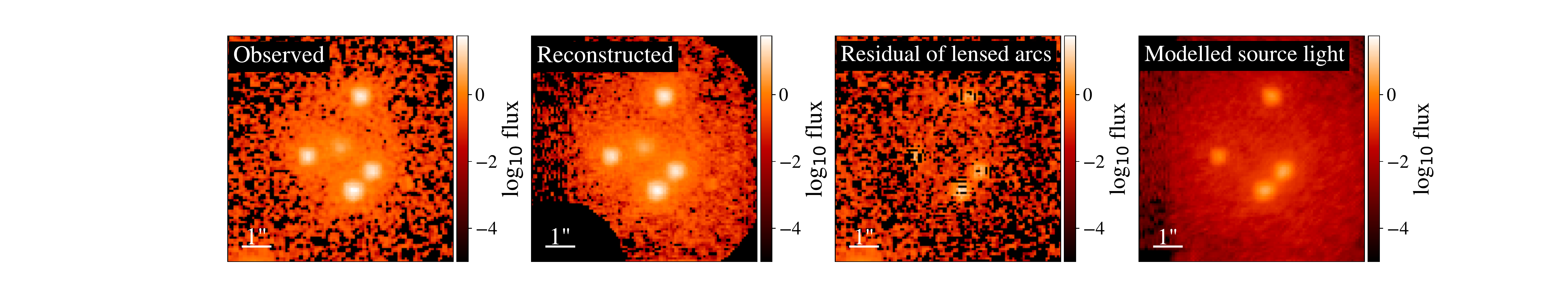}}\\
\subfloat[WFI2033 F140W]{\includegraphics[trim = 75mm 12mm 45mm 12mm, clip, width=0.9\textwidth]{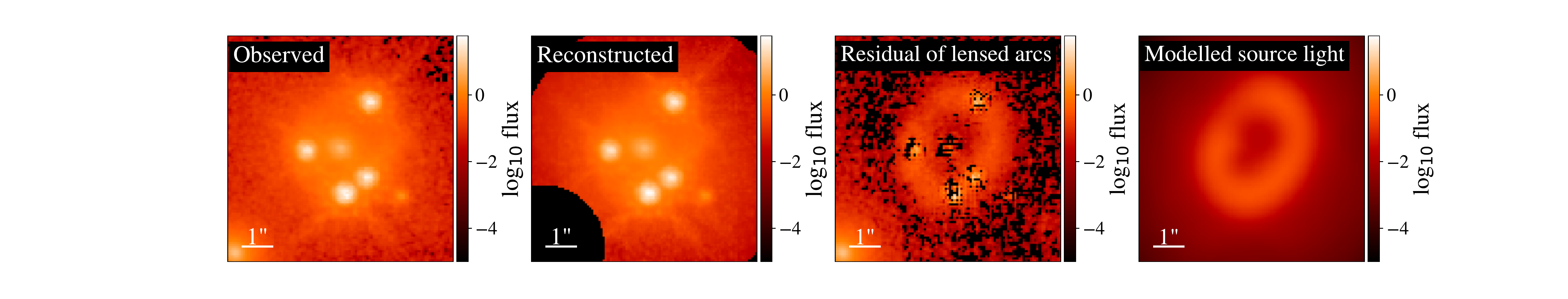}}\\
\subfloat[WFI2033 F555W]{\includegraphics[trim = 75mm 12mm 45mm 12mm, clip, width=0.9\textwidth]{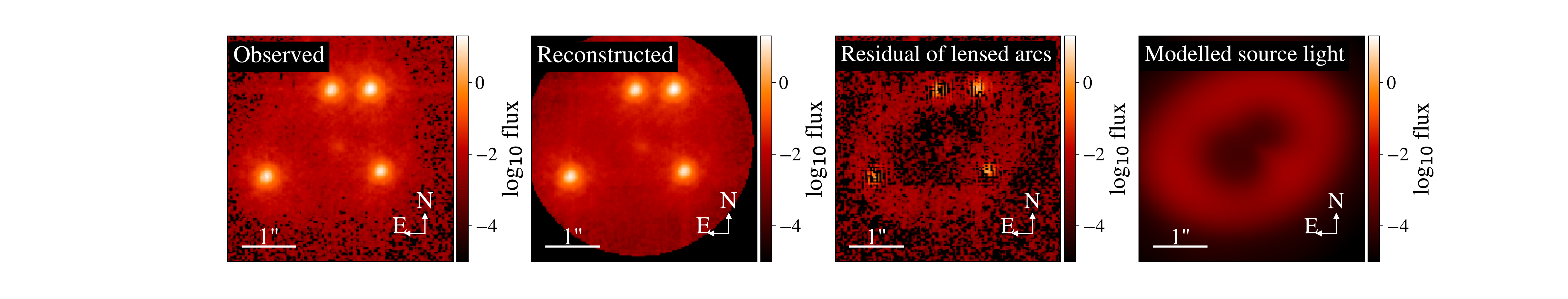}}\\
\subfloat[WFI2033 F814W]{\includegraphics[trim = 75mm 12mm 45mm 12mm, clip, width=0.9\textwidth]{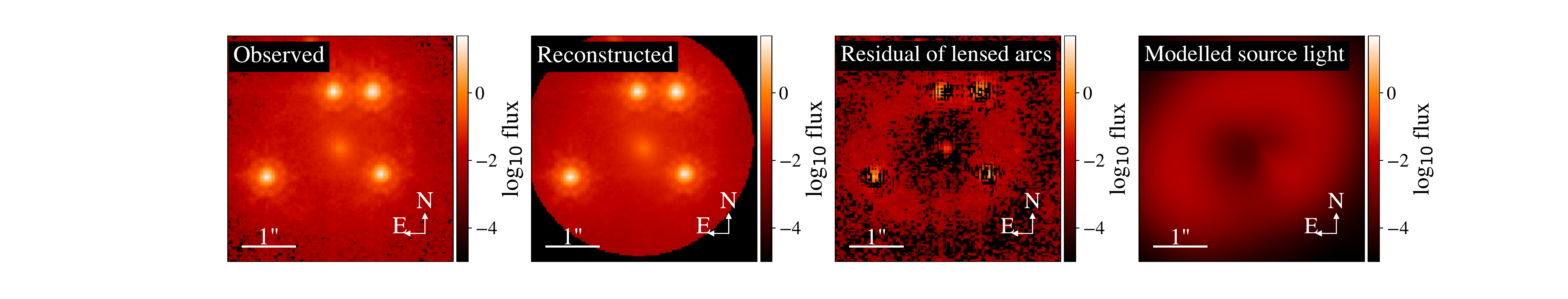}}\\
\subfloat[Host galaxy stellar template inference]{\includegraphics[trim = 10mm 0mm 20mm 15mm, clip, width=0.6\textwidth]{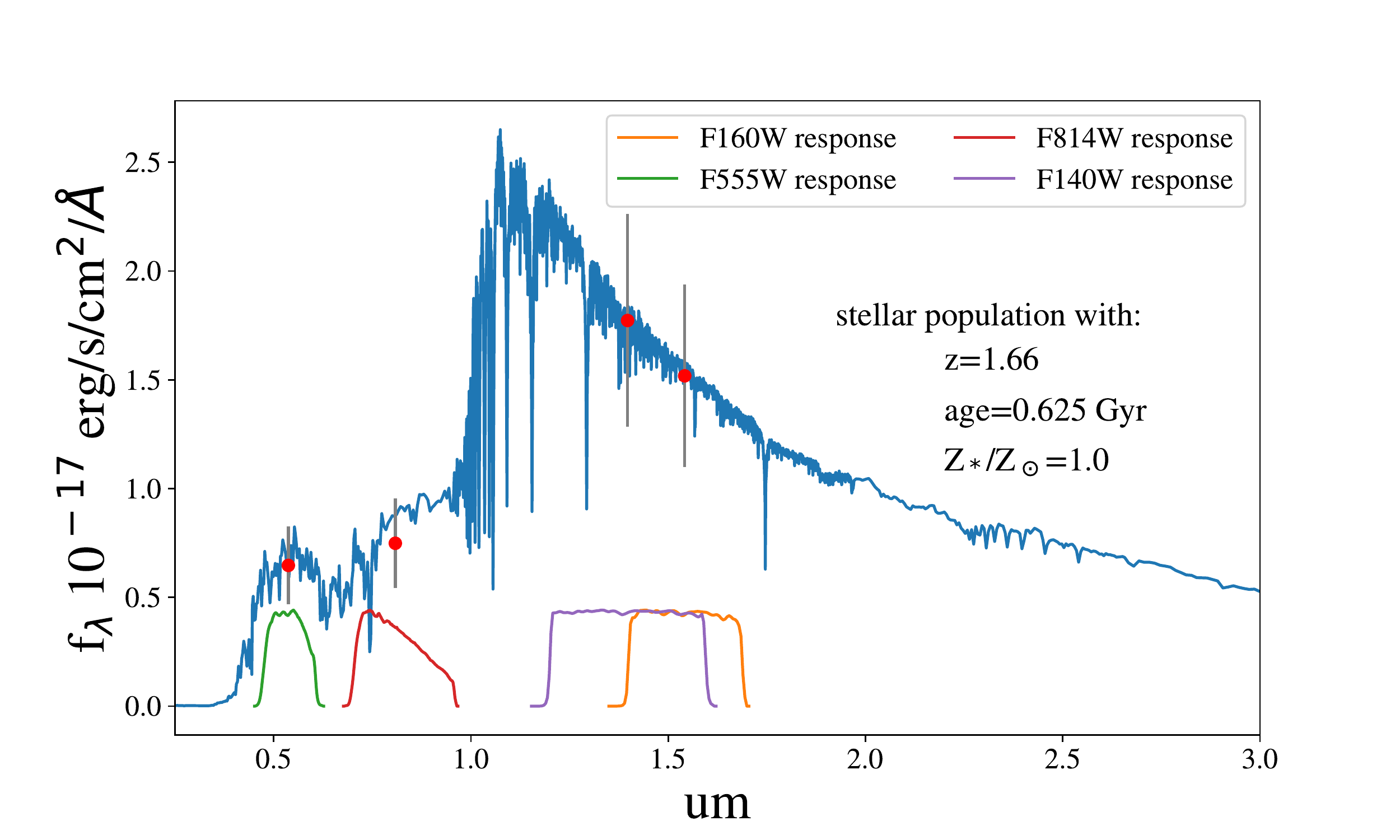}}
\caption{\label{fig:app_WFI2033} 
Same as Figure~\ref{fig:app_WFI2033} but for WFI2033. The F125W dare are too shallow to detect the host and are thus omitted in the fit.}
\end{figure*}

{\bf WFI2033} ~ 
%In addition to the F160W filter, images through the four filters F125W, F140W, F555W, F814W are also available in the \hst\ archive. As for HE0435, we infer the host color in the image plane. We find a stellar population with 0.625 Gyr is a good match to the colors.
In addition to WFC3/F160W, WFI2033 has been observed by \hst\ in  four bands, i.e., WFC3/F125W (GO-12874; PI: D. Floyd), WFC3/F140W (GO-13732; PI: A. Nierenberg), ACS/F555W+F814W  (GO-9744; PI: C. S. Kochanek). Similar to HE0435, the lens modelling and the host photometry of WFI2033 have been inferred through these four bands. The results are presented in Figure~\ref{fig:app_WFI2033}-(a)$-$(d). Unfortunately, the  F125W data are too shallow to robustly detect the host.  Thus, we do not use the F125W band for the SED fitting. Using four-band photometry, we find a stellar population of 0.625 Gyr age provides a good match to the colors, as shown in Figure~\ref{fig:app_WFI2033}-(e).

{\bf HE1104 and the other systems} ~ HE1104 has also been observed by \hst\ through the F555W and F814W bands. However, given the limited exposure time in these two bands, the lensed arcs are too faint to be detected; thus, we are not able to infer the color of the host for HE1104. No multi-band information is available for the other systems. Thus, for these cases, we follow D20 and assume a typical stellar population age, i.e., 1 Gyr and 0.625 Gyr for systems as $z<1.44$ and $z>1.44$, respectively. Of course, this choice is not unique. However, we stress that as we are mostly interested in comparison with D20 this strategy is meant to minimize differences between the two approaches.

A summary of the adopted stellar population ages is given in Table~\ref{tab:host_measure}, column~(7). Applying these templates to the filter magnitudes obtained in last section, we derive the stellar mass of our system, Table~\ref{tab:host_measure}, column~(8).

Considering the simulations by~\citet{Ding2017a} and the fact that we are able to obtain very consistent host magnitudes for the HE0435, RXJ1131 and WFI2033 using the independent approaches, the fidelity of the inferred magnitude is expected to be well characterized by the quoted uncertainties.
Our systems do not have sufficient multi-band information to constrain their metallicity and star formation histories. Therefore, we adopt a simple stellar population template to calculate \mstar, but reflect the lack of information in the associated uncertainty. Following \citet{Bell2001}, we estimate a typical uncertainty of 0.2~dex for all the \mstar\ measurements in this paper.

\subsection{The \mbh-\mstar\ relation}\label{sec:relation}
\begin{figure*}
\centering
\begin{tabular}{c c}
{\includegraphics[height=0.45\textwidth]{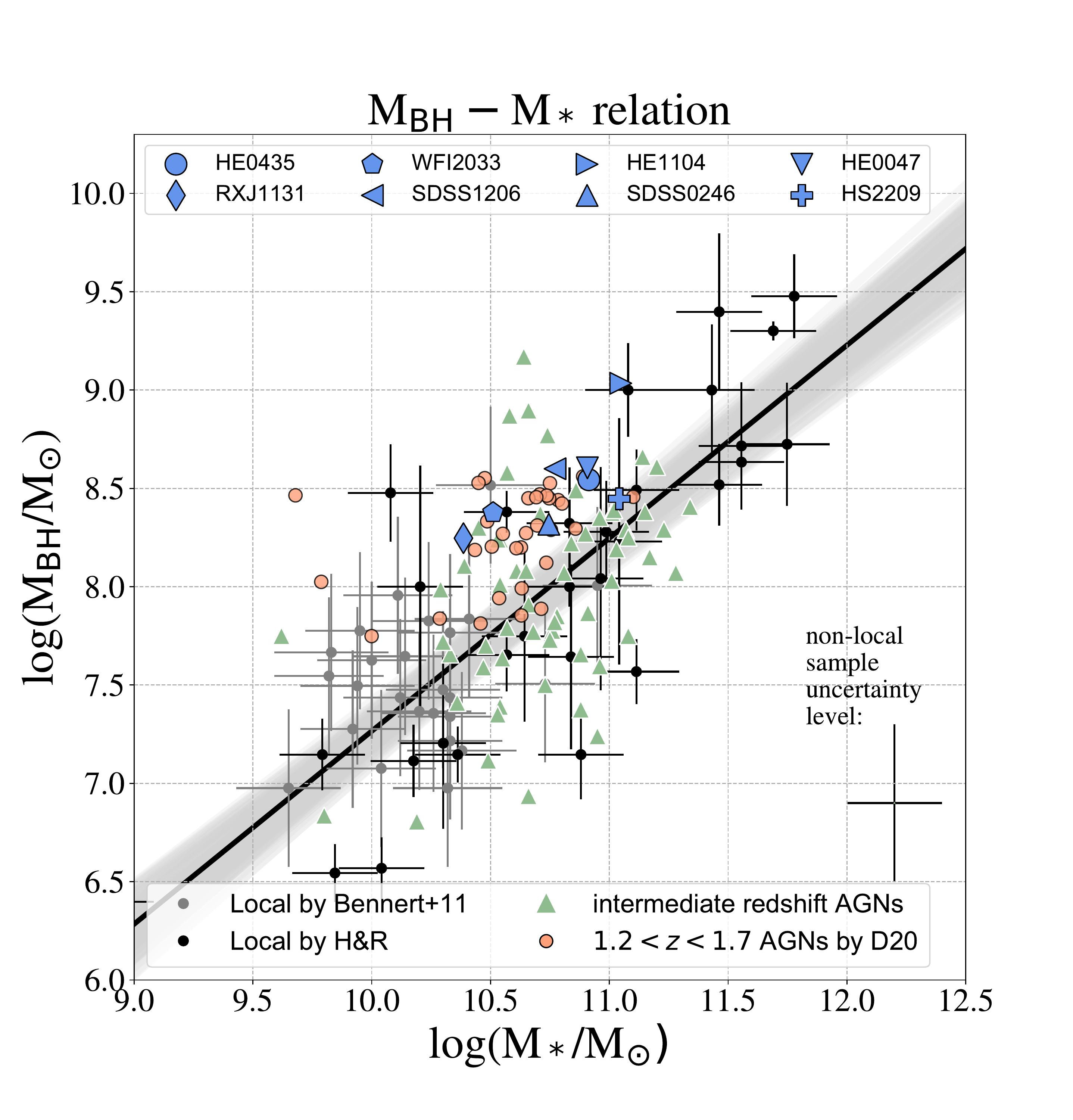}}&
{\includegraphics[height=0.45\textwidth]{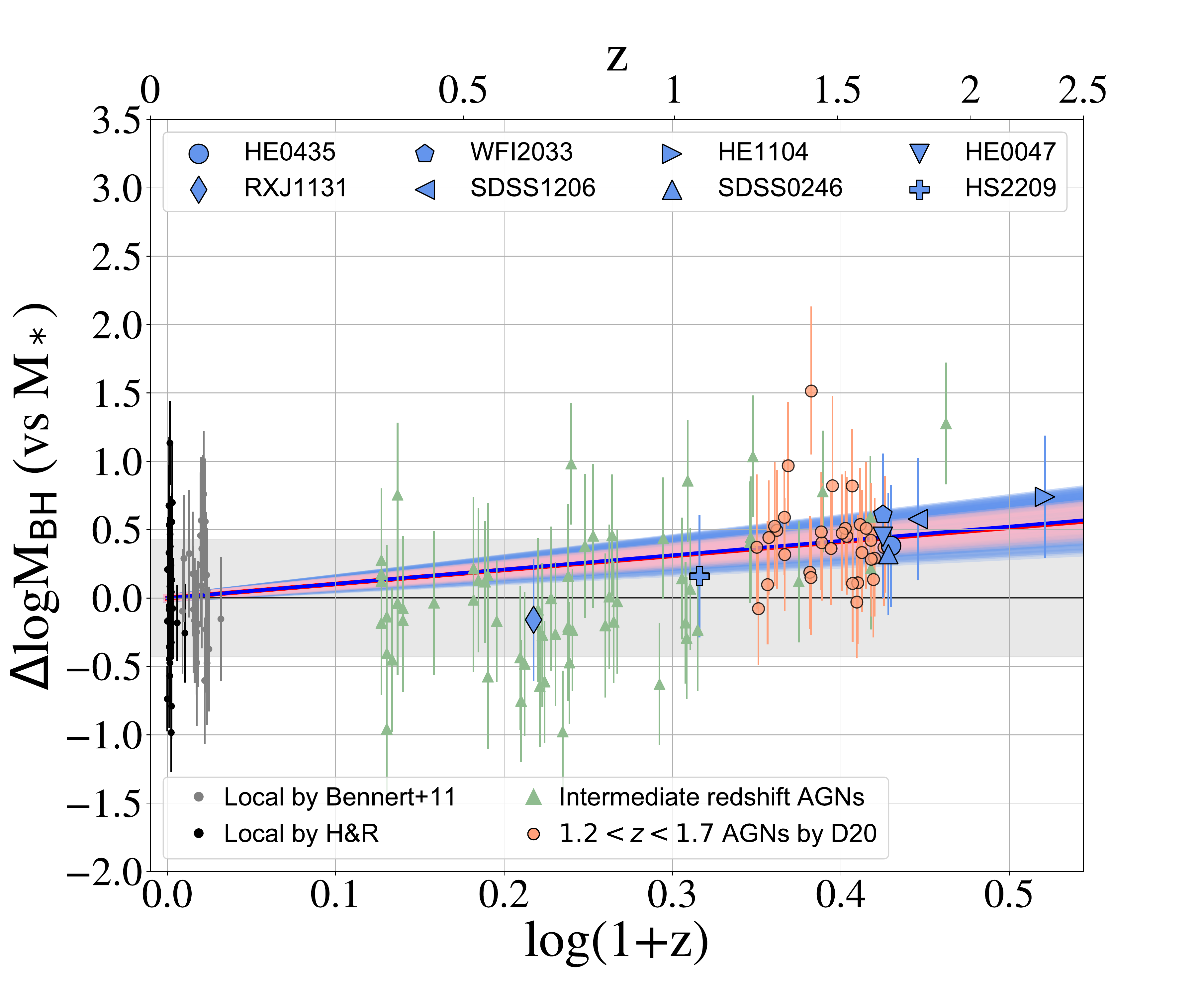}}\\
\end{tabular}
\caption{\label{fig:scaling_relation} 
(left): BH masses vs. stellar mass correlation (\mbh-\mstar). The black line and the gray shaded region indicate the best-fit and 1$\sigma$ confidence interval of the form given by Equation~\ref{eq:MMlocal}.
(right): Offset of $\log(\mathcal M_{\rm BH})$ at fixed \mstar as a function of redshift. Our new measurements are overlaid on samples of non-lensed AGNs taken from the literature and measured in a self consistent way to facilitate a direct comparison. The red line and shaded region are the best-fit and $1-\sigma$ results of fitting  Equation~\eqref{eq:offset} to the 32 high-z AGNs measured by D20. The blue line and region are the results based on the eight lensed systems presented in this work. The blue and red areas are in agreement within the errors.}
\end{figure*}

In Figure~\ref{fig:scaling_relation}-(left), we plot \mbh\ vs \mstar\ for our sample, together with the comparison sample introduced in Section~\ref{sec:sample_select}.  We expect that the uncertainty of the \mbh\ (i.e., $0.4~$dex) dominates the error budget for the entire sample. Following D20, we adopt, as a baseline, a local \mbh-\mstar\ of the form:
\begin{eqnarray}
\label{eq:MMlocal}
\log \big( \frac{\mathcal M_{\rm BH}}{10^{7}M_{\odot}})= \alpha + \beta \log(\frac{M_*}{10^{10}M_{\odot}}),
  \end {eqnarray}
with  $\alpha = 0.27$, $\beta = 0.98$ based on the local sample of 55 objects measured in a consistent manner. We find that our lensed systems are above the local relation, consistent with the inference from the 32  non-lensed AGNs published by D20 in a  similar redshift range. To quantify the evolution as a function of redshift, we parameterize it as:
\begin{eqnarray}
\label{eq:offset}
\Delta \log \mathcal M_{\rm BH}= \gamma \log (1 + z),
\end{eqnarray} 
where $\Delta \log \mathcal M_{\rm BH}$ is the offset of \mbh\ with respect to the local baseline at fixed \mstar. To make a direct comparison, we reproduce the plot shown in Figure~8 of D20. Then, we add our new lensed AGN measurements and show the result in Figure~\ref{fig:scaling_relation}-(right). We find that the offset from the local relationship is similar for the two samples. Based on Equation~\eqref{eq:offset}, we fit the evolution of the eight lensed systems and obtain $\gamma=1.05\pm0.44$, which agrees with the results of D20 ($\gamma=1.03\pm0.25$ using 32 AGN) within $1-\sigma$ level. The consistency between the two measurements provides an important verification of the accuracy of the results and strengthens the conclusions drawn by D20 that the {\it observed} value of \mbh\ at a fixed \mstar\ tends to be larger at higher redshift than the ones in the local universe. It is possible that the uncertainty of the \mstar\ could be higher than our assumed 0.2 dex, given that the star formation histories could vary widely at high redshift ($z\sim$2). However, increasing the uncertainty on the stellar mass does not change significantly our conclusions, since the error budget is dominated by the uncertainty in black hole mass. For example, increasing the stellar mass uncertainty to $0.3$ ($0.4$)~dex, we find $\gamma = 1.05\pm0.48$ $(1.03\pm0.53)$.

Note that this apparent evolution is obtained directly using the observed sample, before considering selection effects~\citep{Tre++07,Schulze2011,Bennert++2011, Schulze2014,Park15}. For instance, adopting the framework introduced in~\citet{Schulze2015}, D20 estimated that accounting for selection biases would yield a more modest evolution $\gamma=0.50\pm0.25$. The lensed AGN systems considered in this work were selected for time-delay cosmography, based on the availability of a time delay and the known detectability of the host galaxy. This is a complex selection function, different than the one used by D20. Although it is encouraging that the two samples present the same apparent evolution, inferring the true underlying evolution requires modelling the selection function, as studied by D20. A full modelling of the complex selection function is not warranted given the small size of the lensed quasars sample. Thus, at this stage, the lensed quasars should be considered as a check on possible systematic measurement error in the D20 analysis rather than a stand-alone measurement.

We consider all random and systematic effects in Appendix~\ref{sec:diss}. To summarize, the uncertainty level of the \mstar\ in this work is assumed as 0.2~dex  as mentioned in the previous section. The uncertainty in the black hole mass estimates is at a level of 0.4~dex, which dominates the overall error budget.

\section{Conclusions}\label{sec:con}
Using eight strongly lensed AGN systems, we presented new measurements of the correlations of the mass of supermassive black hole with the stellar masses of their host galaxy. We adopt state-of-the-art lens modelling techniques to estimate the magnitude host galaxy, in terms of a standard \sersic\ profile. We estimated the \mbh\ of our sample using a set of calibrated single-epoch estimators to assure self-consistency and consistency with the comparison samples taken from the literature.

We directly compare our sample to the recent measurements by~\citet[][D20]{Ding2020a}, who used the same approach to derive the \mstar\ and calibrated the \mbh\ using consistent recipes. The \mbh-\mstar\ correlation and its evolution with cosmic time are in excellent agreement with the results obtained by D20, as shown in Figure~\ref{fig:scaling_relation}. Currently, the sample of lensed AGN (i.e., eight systems) is quite limited in size; as a result, given the statistical errors it is not worth modelling selection effect. Nevertheless, the good agreement with the D20 sample strengthens their conclusions, which can be summarized as follows. First, the growth of the supermassive black hole somewhat predates that of its host galaxy during their co-evolution, even when considering total stellar mass, as it is often done at high redshift. However, the actual morphologies of these hosts are likely to be more complex, including a disk and bulge component, as in the case of  RXJ1131. In contrast, the galaxies in the local sample are typically bulge-dominated and the bulge mass is adopted as their  \mstar. Thus, the reported evolution is weaker than what would be inferred by comparing \mbh\ to bulge \mstar\ at all redshifts \citep{Bennert++2011}. Taken together, these results are consistent with a scenario in which the stellar mass is transferred from disk to the bulge at a faster rate than the growth of \mbh\ since $z\sim2$.

Our work based on highly magnified AGN showcases the power of strong lensing to effectively increase the resolution of a telescope and shows that uncertainties related to lens modelling are subdominant with respect to other sources of uncertainty like black hole mass. Furthermore, our work provides a powerful verification of the fidelity of the host galaxy reconstruction in non-lensed AGN. 

%\pink{Strong lensing can also help to extend the measurements of the AGN hosts to higher redshift.}

In conclusion, lensed AGNs have great potential to extend the study of the \mbh-\mstar\ correlation to higher redshifts than those considered here.  So far, this kind of work has been limited by sample size. However, given the pace of discovery of lensed quasars in imaging and spectroscopic surveys ~\citep[e.g.,][]{Oguri2010, Agn++15,Mor++16,Sch++16,Ost++17,Tre++18, Ang++18, Lem++20}, the samples of lensed AGNs with hosts that can be recovered with high fidelity is likely to continue to grow in wide field imaging and spectroscopic surveys. The forthcoming launch of the {\it James Webb Space Telescope} and the first light of adaptive optics-assisted extremely large telescopes may provide high-quality imaging data of AGNs up to the highest redshift at which they have been discovered.

\section*{Acknowledgements}
The authors thank the anonymous referee for helpful suggestions and comments which improved this paper.
We are grateful to Frederic Courbin, Leon Koopmans for useful comments
and suggestions that improved this manuscript. We thank John Silverman and Vardha Bennert for many conversations on the topic of galaxy and black hole co-evolution.

This work is based in part on observations made with the NASA/ESA Hubble Space Telescope, obtained at the Space Telescope Science Institute, which is operated by the Association of Universities for Research in Astronomy, Inc., under NASA contract NAS 5-26555. X.D., S.B., and T.T. acknowledge support by the Packard Foundation through a Packard Research fellowship to T.T. This project is received support by the NSF through grant 1907208. This work is supported by JSPS KAKENHI Grant Number JP18H01251 and the World Premier International Research Center Initiative (WPI), MEXT, Japan. This project has received funding from the European Research Council (ERC) under the European Union's Horizon 2020 research and innovation programme (grant agreement No 787886).
C.D.F. acknowledge support for this work from the National Science Foundation under Grant Numbers AST-1312329 and AST-1907396.
SHS thanks the Max Planck Society for support through the Max Planck Research Group.

This work has made use of \lenstronomy~\citep{lenstronomy}, {\sc Astropy}~\citep{Astropy}, {\sc photutils}~\citep{photutils}, {\sc Matplotlib}~\citep{Matplotlib} and standard Python libraries.

\section*{Data Availability}
The data underlying this article are available in the article and in its online supplementary material.

%%%%%%%%%%%%%%%%%%%%%%%%%%%%%%%%%%%%%%%%%%%%%%%%%%



\bibliographystyle{mnras}
%\bibliography{reference} % if your bibtex file is called example.bib
\input{manuscript.bbl}

%\begin{thebibliography}{99}
%\end{thebibliography}
%%%%%%%%%%%%%%%%%%%%%%%%%%%%%%%%%%%%%%%%%%%%%%%%%%

%%%%%%%%%%%%%%%%% APPENDICES %%%%%%%%%%%%%%%%%%%%%

\appendix

\section{Photometry Inference}\label{sec:photometry}
We describe the details of the fitting for each system and present the inference of the photometry of the host galaxy.

\subsection{HE0435}\label{subsec:HE0435}
We follow the approach described in Section~\ref{sec:Modelling}. We select 5 isolated stars in this field as initial PSFs to input to the fitting, for a total of 60 fits. Based on the top-eight choices, we perform the weighting algorithm and measure the host flux, host-to-total flux ratio, effective radius and \sersic\ index. The inference results are shown in Table~\ref{tab:host_measure}, (2)$-$(6) columns.

The inference by the best-fit lens model for HE0435 is shown in Figure~\ref{fig:image_inference}-(a). Not surprisingly, we note that the residuals level in the normalized plot appears to be larger than the ones presented by~\citet{Wong2017}. This is primarily due to the fact that the surface brightness of the host galaxy in our model is defined as a \sersic\ profile, which is relatively simple and smooth compared to the pixellated reconstruction technique adopted by {\sc Glee}~\citep{Suy++06, Halk2008, S+H10, Suyu2012} or the shapelet technique adopted by \lenstronomy~\citep{Birrer2015}. The smooth features of \sersic\ profile can not capture the clumps in the host galaxy, such as the star-forming regions. Nevertheless, our approach is sufficient to derive a self-consistent one-step inference of the global host light in terms of the \sersic\ flux, i.e., the quantity commonly measured in the literature for non-lensed samples. 

As a cross-check, we compare our host magnitude to the inference of HE0435 by~\citet{Ding2017b}. They inferred the \sersic\ magnitude by fitting to the pixelized host galaxy as reconstructed by~\citet{Wong2017}. The host magnitude measured by~\citet{Ding2017b} is: $mag = 21.75 \pm 0.13$. %, $R_{eff} = 0.82 \pm 0.14$ and $n = 3.94 \pm 0.14$. 
The results are in excellent agreement with the measurement reported here, i.e., $mag = 21.50 \pm 0.35$.

\subsection{RXJ1131}
A lens model of RXJ1131 based on ACS/F814W data has been presented by~\citet{Suyu2013}. The host galaxy of this system is lensed to an extended arc. A clear bulge and disk component can be identified. Thus, we describe the host galaxy with two \sersic\ profiles with index values $n$ fixed to $1$ and $4$, respectively, to mimic the light distribution of the disk and bulge. In addition, following~\citet{Suyu2013}, we consider the perturbations by the small object ($0\farcs{5}$ in the north) and describe it as a SIS and \sersic\ for its mass and light, respectively.

The photometry of the RXJ1131 host galaxy is fitted with a set of 4 initial PSFs. The results are summarized in Table~\ref{tab:host_measure}, with the best-fit result shown in Figure~\ref{fig:image_inference}-(b).

We also compare our measurement to the previous reconstructions by \citet{Ding2017b}. Based on the reconstructions by~\citet{Suyu2013}, the inference of the host magnitudes by~\citet{Ding2017b} are $mag_{\rm bulge} = 21.81 \pm 0.28$ and $mag_{\rm disk} = 20.07\pm0.06$. The results are in excellent agreement with our inferred bulge magnitude ($21.80\pm0.21$)   shown in Table~\ref{tab:host_measure}. However, our inferred disk magnitude ($19.33\pm0.16$) is brighter than that reported by \citet{Ding2017b}. This difference is not surprising because our reconstruction of the host galaxy is based on a much more extended region to collect the disk light, compared with~\citet{Suyu2013} who performed the lens modelling using a smaller lens mask (see Figure~4 therein). Note that, at variance with the procedure described here, \citet{Ding2017b} used a \sersic\ to fit the pixellated source reconstructed by~\citet{Suyu2013} in the source plane. We checked that finite grid effects did not introduce any substantial difference. As a sanity check, we find that our inferred effective radius is very consistent, $0\farcs{}90 \pm 0\farcs{}06$ (this paper) and $0\farcs{}84 \pm 0\farcs09$~\citep{Ding2017b}. The difference between the \citet{Suyu2013} reconstruction and the one presented here makes sense in terms of the different goals of the two studies. While our primary aim is to reconstruct the host galaxy photometry, for \citet{Suyu2013} it was only a byproduct on the way to time-delay cosmography.

\subsection{WFI2033}
WFI2033 is the last quadruply lensed system in our sample whose lens model has been previously investigated \citep{Rusu2019}. There is a satellite galaxy in the north of the lens. However, the satellite galaxy has a much smaller mass than the main deflector. In addition, there is a galaxy west of the main target, which also has a small effect on the total macro-magnification $(<10\%)$. Thus, we ignore their influence on the magnification but only fit the light of the satellite galaxy using a \sersic\ model. 
We select a total of 8 initial PSF stars to model this system.

The final inference results are presented in Table~\ref{tab:host_measure} and Figure~\ref{fig:image_inference}-(c). We compare our inference to the previously reconstructed host galaxies. Modelling the reconstructed host by~\citet{Rusu2019}, (i.e., the Figure~4 bottom-right plane therein) as a \sersic\ profile, we infer the $mag = 21.98 \pm 0.15$, which is very consistent to our inference (i.e., $mag = 21.78 \pm 0.25$).

\subsection{SDSS1206}
SDSS1206 is a unique system -- the AGN is doubly imaged by the deflector while most of the host falls inside the inner caustic and ends up being quadruply imaged. Following~\citet{Birrer2019}, we consider the galaxy triplet group at the north-west and use a single SIS model to denote their overall mass perturbation. Moreover, as noted by~\citet{Birrer2019}, a sub-clump is located in the north which is hardly visible \citep[see Figure~1 in][]{Birrer2019}. We model this sub-clump as a SIS mass model and a circular \sersic\ light model with joint centroids. It is worth noting that we are using the same imaging modelling tool as~\citet{Birrer2019}.

Due to the limited number of stars in the field of view, there are only 2 stars available as initial PSF. To expand the volume of modelling options, we also take the stack of the two bright stars as derived by~\citet{Birrer2019}, as a third initial PSF. We find visible residuals at the fitted lensed arcs region using a single \sersic\ model as host. However, a double \sersic\ model does not significantly improve the goodness of fit. Thus, we adopt the single \sersic\ model in our final inference. Our inferred results are presented in Table~\ref{tab:host_measure} and Figure~\ref{fig:image_inference}-(d).

\subsection{HE1104}
HE1104 is a typical doubly imaged quasar. We have selected in total of 5 initial PSFs to perform the fit. There is an object in the northeast. However, since we do not know its redshift and considering it is further away from the lens we do not model it explicitly but just mask it out in the fitting.

The inference is presented in Table~\ref{tab:host_measure} and Figure~\ref{fig:image_inference}-(e). the lensed arcs can be clearly seen from the bottom-left panel, indicating that the host galaxy is well detected.

It is worth noting that the HE1104 has been modelled by~\citet{Peng2006} based on the \hst/NICMOS H-band (F160W) imaging data. Their inferred host light is $20.14\pm0.30$ mag in Vega system, which is also consistent to our inference ($20.00\pm0.15$ mag in Vega).
%zeropoint 25.9463(AB) - 24.6949(Vega) = 1.2514
%HE1104 mag:  21.25 (AB)  --> 20.00 (Vega)

\subsection{SDSS0246}
Having been imaged with WFC3-UVIS/F814W, the resolution of the data for this system (together with the remaining two systems) is much higher than for those imaged in the IR, with a drizzled pixel scale of $0\farcs{}03$. However, the arcs are much fainter compared with those of other systems imaged in the IR band. As a result, fewer pixels with signal are available for the fit than for other systems. Nevertheless, the host inference is successfully reconstructed as shown in Figure~\ref{fig:image_inference}-(f) and Table~\ref{tab:host_measure} (3 initial PSF guesses were adopted).  

\subsection{HS2209}
HS2209 was imaged by \hst\ during two visits ($vis05$ and $vis06$) at different orientations. We modelled the two visits separately and recovered mutually consistent host galaxy magnitudes. However, we found that the data from {\it vis06} can be modelled with smaller residuals; thus the inference based on {\it vis06} was adopted as our best estimate (using 7 initial PSF stars), as listed in Table~\ref{tab:host_measure}. The inference for this system is summarized in Figure~\ref{fig:image_inference}-(g).

\subsection{HE0047}
\begin{figure*}
\centering
\begin{tabular}{c c}
\subfloat[HE0435]{\includegraphics[trim = 60mm 25mm 20mm 25mm, clip, width=0.5\textwidth]{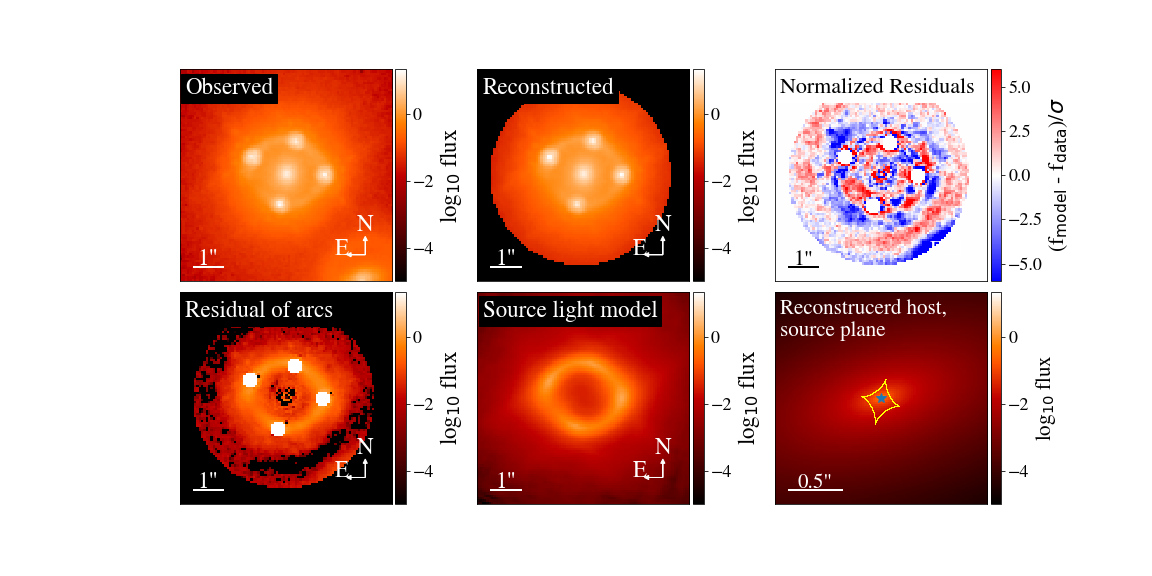}}&
\subfloat[RXJ1131]{\includegraphics[trim = 60mm 25mm 20mm 25mm, clip, width=0.5\textwidth]{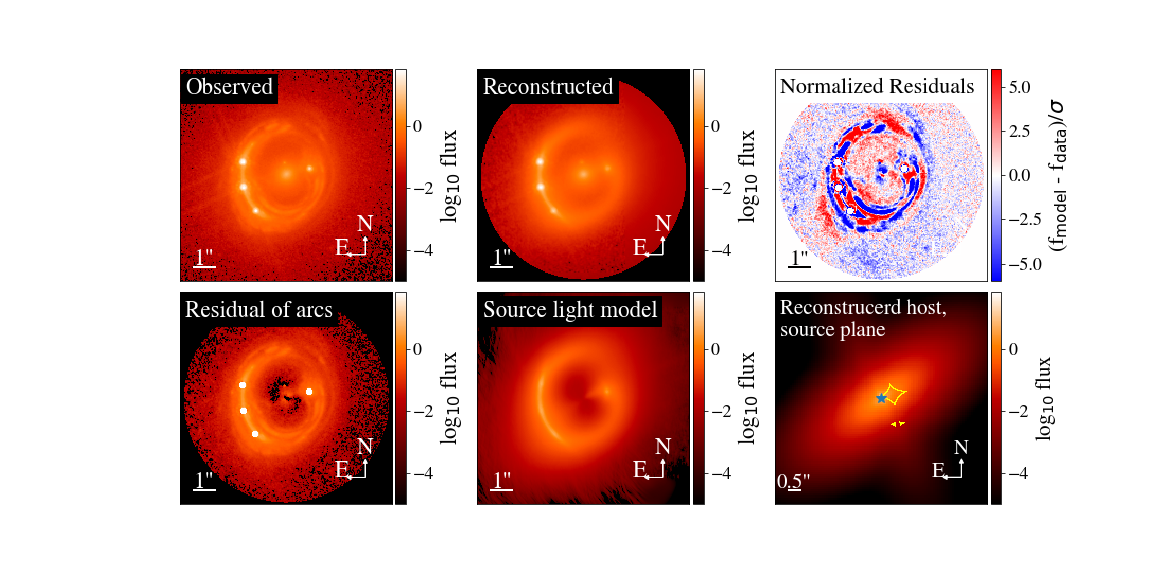}}\\
\subfloat[WFI2033]{\includegraphics[trim = 60mm 25mm 20mm 25mm, clip, width=0.5\textwidth]{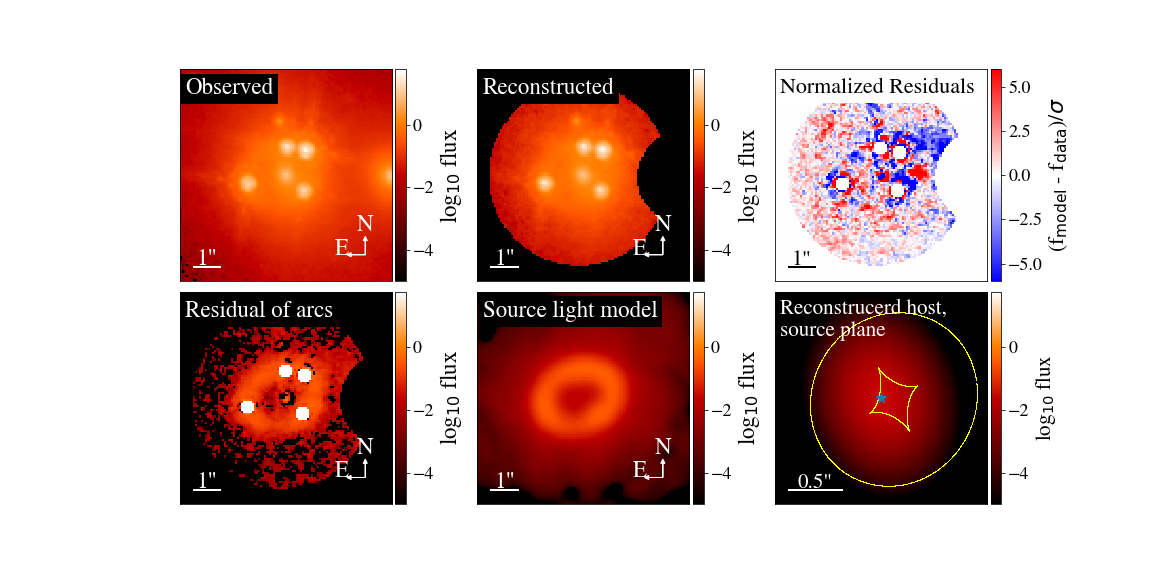}}&
\subfloat[SDSS1206]{\includegraphics[trim = 60mm 25mm 20mm 25mm, clip, width=0.5\textwidth]{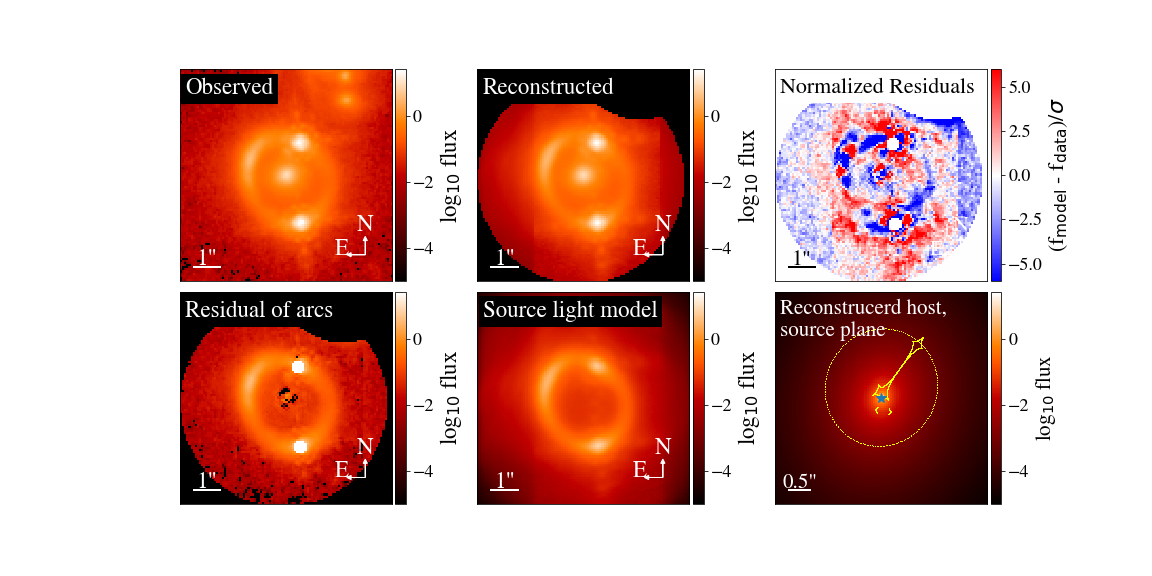}}\\
\subfloat[HE1104]{\includegraphics[trim = 60mm 25mm 20mm 25mm, clip, width=0.5\textwidth]{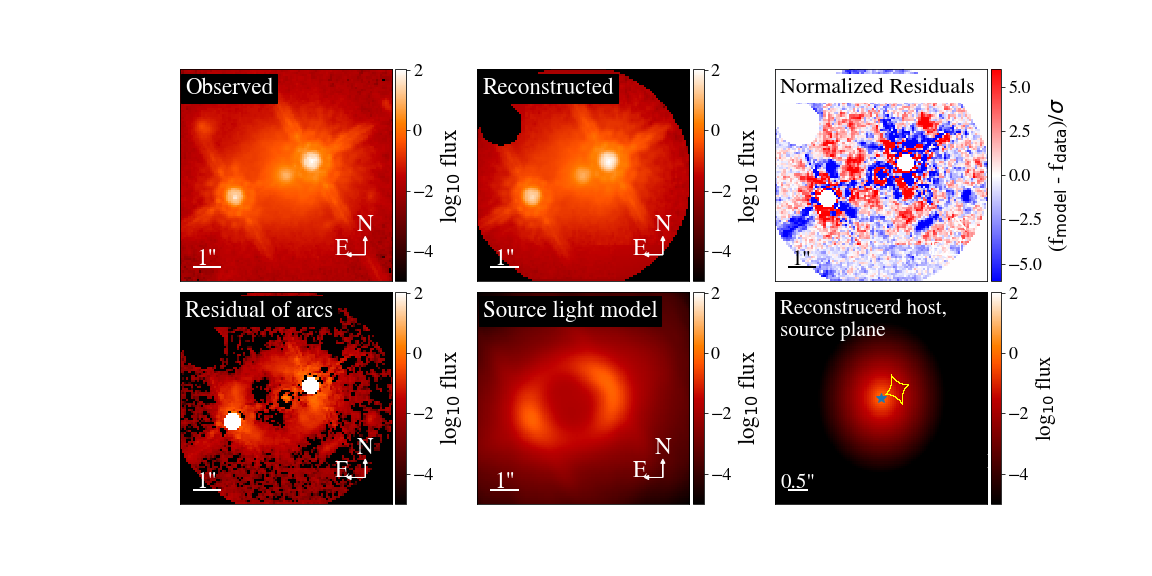}}&
\subfloat[SDSS0246]{\includegraphics[trim = 60mm 25mm 20mm 25mm, clip, width=0.5\textwidth]{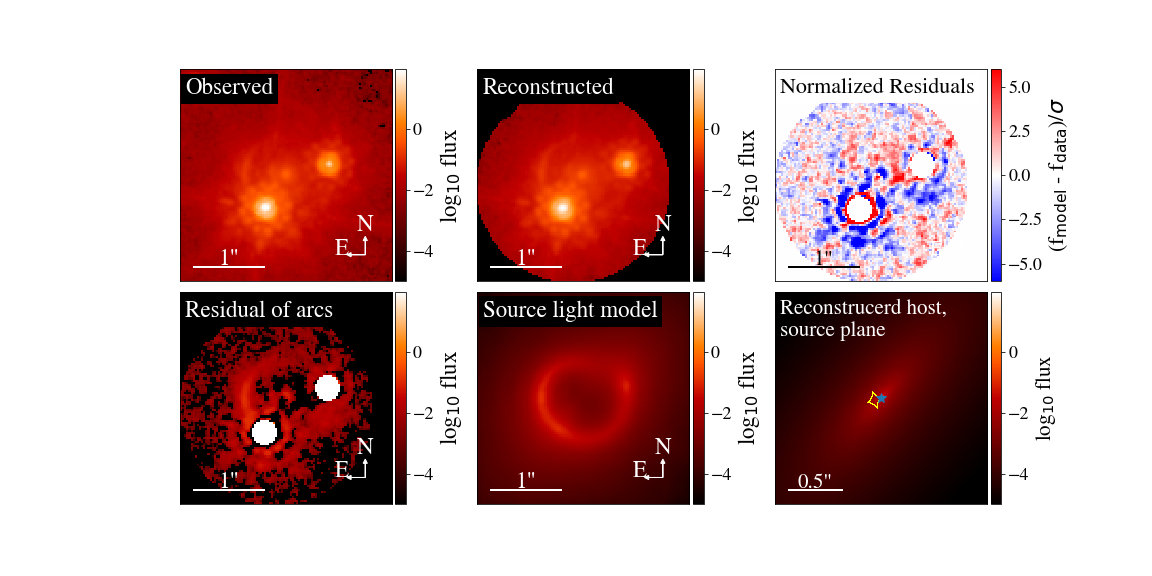}}\\
\subfloat[HS2209]{\includegraphics[trim = 60mm 25mm 20mm 25mm, clip, width=0.5\textwidth]{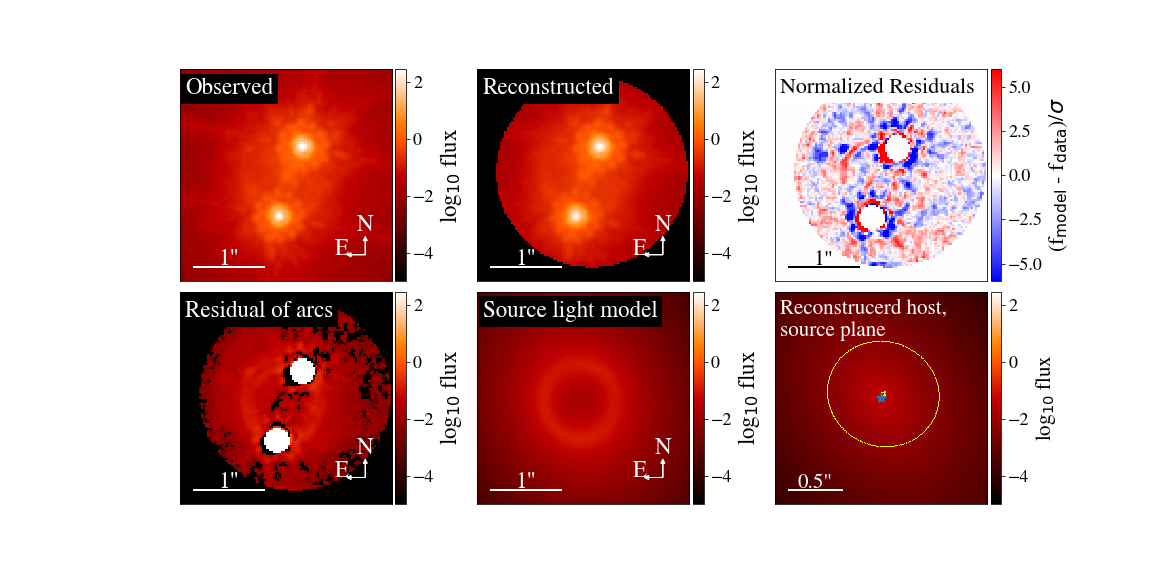}}&
\subfloat[HE0047]{\includegraphics[trim = 60mm 25mm 20mm 25mm, clip, width=0.5\textwidth]{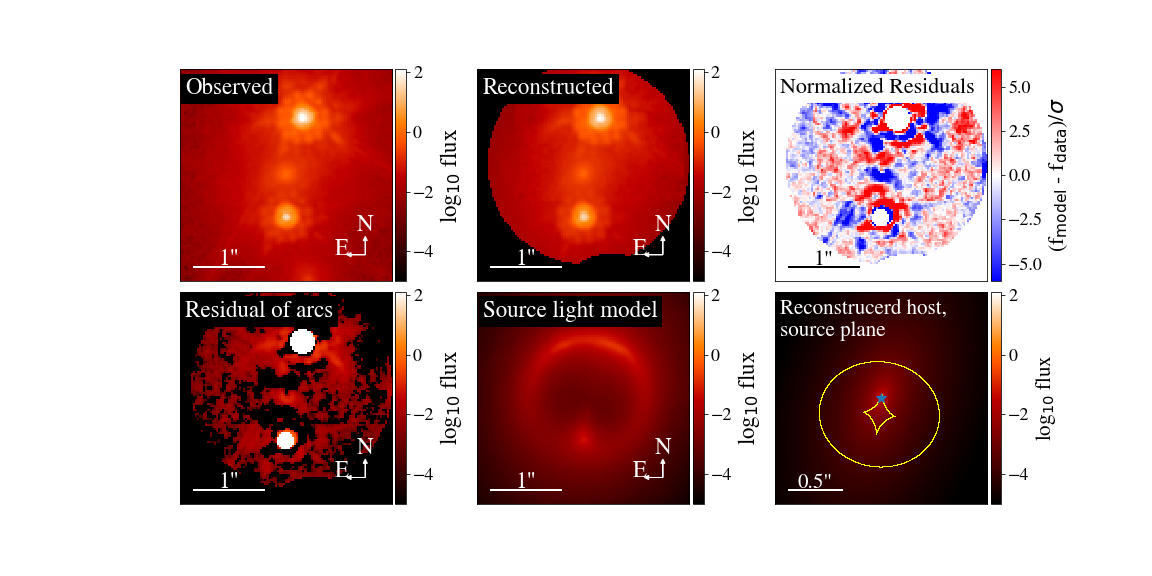}}\\
\end{tabular}
\caption{\label{fig:image_inference} 
Illustrations of the inference using the best-fit lens model for each system using the AGN center noise level boosted approach. For each figure, the panels from left to right are the follows: Top row: (left to right): (1) observed data, (2) best-fit model (3) residuals divided by the uncertainty level. Some of the residual looks clumpy; however, they do not affect the host property measurements, for more details see Section~\ref{sec:diss} ; Bottom row: (left to right): (4) data minus the model PSF and deflector  (i.e., pure lensed arc image), (5) the model of the lensed arc, (6) reconstructed host galaxy in the source plane with caustic line drawn as yellow line. Across all the systems, the lensed arc feature can be clearly seen in the fourth panel, indicating strong evidence of a detection. In panel (3) and (4), we use white regions to indicate the area where the noise level is boosted. 
}
\end{figure*}

HE0047 is the most challenging system in our sample with the lowest SNR of the lensed arcs. The results, based on 3 initial PSFs, are summarized in Figure~\ref{fig:image_inference}-(f) and Table~\ref{tab:host_measure}. The host magnitude of the HE0047 system has a relatively large uncertainty as reflected in the error bars.

\section{Systematic Errors}\label{sec:diss}

We considered a set of modelling choices to perform the fitting and the final inference is based on a weighting of top-ranked choices. In particular, we treat the two different modelling approaches, i.e., {\it noise boost} and {\it PSF iteration} equally, to derive the averaged results. Of course, the results are somewhat dependent on the weighting scheme, for example, the dispersion of the results by the top-ranked choices. However, the dependency is smaller than other sources of uncertainty. Thus, using different weighting schemes would only change the results marginally ($<0.1$~dex).

We used a range of mass slope values (i.e., 1.9, 2.0, 2.1) to perform the lens modelling. Then, we used our weighting algorithm introduced in Section~\ref{sec:Modelling} to estimate the systematic uncertainty of our inference and assumed it covers the truth. We apply this method to the entire sample to ensure self-consistency within our sample, even though four systems (HE0435, RXJ1131, WFI2033, and SDSS1206) have been analyzed by H0LiCOW collaboration, and have high precision slope measurements available. As a sanity check, we calculate the weighted slope value and make a direct comparison to the H0LiCOW inference. The results are the following (here {\it v.s.} H0LiCOW, the error bars are the $1-\sigma$ level): HE0435~($2.032\pm 0.07$ {\it v.s.} $1.93\pm0.02$), RXJ1131~($2.02\pm 0.06$ {\it v.s.}$1.95\pm0.045$), WFI2033~($1.94\pm 0.07$ {\it v.s.} $1.95\pm0.02$), SDSS1206~($1.98\pm 0.06$ {\it v.s.} $1.95\pm0.05$). The consistency of the results supports the robustness of the systematic uncertainty estimated in this work.

Following standard practice in galaxy evolution studies, we use a \sersic\ model to describe the surface brightness of the host galaxy. The \sersic\ profile is relatively simple and smooth, and cannot capture the clumps in the host galaxy. Thus the smoothness of the \sersic\ profile leads to the relatively large residuals shown in Figure~\ref{fig:image_inference}. However, as mention in Section~\ref{subsec:HE0435}, our goal to derive a self-consistent one-step inference of the host properties to make comparison with the measurements of the non-lensed AGN samples, which are measured using the same methodology. Note that, the methodology has to be consistent between lensed and unlensed AGNs, since the use of a different host model may introduce systematic errors. In fact, it is common to have significant residuals when fitting say a \sersic\ profile to a galaxy (e.g., D20). These residuals of course affect the quality of the fit and could increase the systematics. However, these systematics are much smaller than our target precision of 0.4 dex.  The difference in residuals between this work and the H0LiCOW analysis is once again a reflection of the different purposes of the two studies. Whereas fitting the host surface brightness to the noise level is important to determine the gravitational potential with sufficient precision to infer the Hubble constant, it is not necessary when the goal is to infer the luminosity of the host.

In addition, we adopt simple stellar populations to derive the stellar mass. For 5/8 systems in our sample we did not have color information, and we used instead a fixed age depending on redshift. The lack of color information is reflected in an increase of the uncertainty in the inferred \mstar. It is important to stress once again that the main goal of this work is to provide an independent test of the D20 measurement. Since we are using the same stellar population models, any uncertainty in the models or other stellar population assumptions will cause an absolute change, but those will cancel out when looking at relative consistency between this work and D20.

We do not expect foreground extinction to be significant, because the deflectors are all massive elliptical galaxies. As a sanity check, \citet{1999ApJ...523..617F} and \citet{2008A&A...485..403O} report estimates for HE0435, WFI2033 and HE1104 through filter F160W. For HE1104 the estimated extinction is negative, while for HE0435 and WFI2033 the authors do not report an extinction value because standard extinction laws did not fit the data.
If we focus on the 13 ellipticals from~\citet{1999ApJ...523..617F}, the total median extinction is -0.03, which justifies our choice not to apply any correction.
We interpret the negative values reported in the literature as due to the small effect by dust being overshadowed by chromatic microlensing or variability.

In this work, some assumptions have been made to measure the evolution of \mstar-\mbh. For example, a Chabrier IMF was assumed to measure the stellar mass for all samples. To compare our high redshift measurements with the local ones, we adopted the local sample from~\citet{Bennert++2011, H+R04}, rather than other samples available in the literature. We adopt our own recipes to calibrate the \mbh. Of course, different options would shift the absolute value of the inferred \mstar\ and \mbh. However, since the entire sample is self-consistent, a different assumption would only shift the global \mstar-\mbh\ together, leaving the offset value and the evolution conclusion the same. More details can be found in D20, Section 6.

%In summary, considering all random and systematic effects, the uncertainty of \mstar\ relative to that measured by D20 is smaller than the estimated uncertainty in the black hole mass (0.4~dex), which dominates the overall error budget.

%%%%%%%%%%%%%%%%%%%%%%%%%%%%%%%%%%%%%%%%%%%%%%%%%%

% Don't change these lines
\bsp	% typesetting comment
\label{lastpage}
\end{document}